\documentclass[apl,showpacs,byrevtex,floatfix,twocolumn,superscriptaddress]{revtex4}
\usepackage{amssymb,graphicx,afterpage}
\begin{document}

\title{\boldmath New type of ellipsometry in infrared spectroscopy: The double-reference method \unboldmath}
%
%
\author{I. K\'ezsm\'arki}
\affiliation{Department of Physics, Budapest University of
Technology and Economics and Condensed Matter Research Group of the Hungarian Academy of Sciences, 1111 Budapest, Hungary}
\author{S. Bord\'acs}
\affiliation{Department of Physics, Budapest University of
Technology and Economics and Condensed Matter Research Group of the Hungarian Academy of Sciences, 1111 Budapest, Hungary}

\date{\today}
%
%
\pacs{\ }
\begin{abstract}
We have developed a conceptually new type of ellipsometry which allows the determination of the complex
refractive index by simultaneously measuring the unpolarized normal-incidence reflectivity relative to the
vacuum and to another reference media. From these two quantities the complex optical response can be
directly obtained without Kramers-Kronig transformation. Due to its transparency and large refractive index over a broad range of the spectrum, from the far-infrared to the soft ultraviolet region, diamond can be ideally used as a second reference. The experimental arrangement is rather simple compared to other ellipsometric techniques.
\end{abstract}
\maketitle
%

Determination of the complex dielectric response of a material is an everlasting problem in optical
spectroscopy. Depending on the basic optical properties, whether the sample is transparent
or has strong absorption in the photon-energy range of interest, its absolute reflectivity or the transmittance is usually detected with normal incidence. Both quantities are related to the intensity of the
light and give no information about the phase change during either reflection or transmission. Consequently, the
phase shift is generally determined by Kramers-Kronig (KK) transformation in order to obtain the complex
dielectric response. However, for the proper KK analysis the reflectivity or the transmittance spectrum has to
be measured in a broad energy range, ideally over the whole electromagnetic spectrum.

On the other hand, there exist ellipsometric methods \cite{Azzam77,Roseler90} which are capable to
simultaneously detect both the intensity and the phase of the light reflected back or transmitted through a
media. Among them the most state-of-the-art technique is the time domain spectroscopy but its applicability is
mostly restricted to the far infrared region.\cite{Bartels06,Watanabe07} An other class of ellipsometric
techniques, sufficient for broadband spectroscopy, requires polarization-selective detection of
light.\cite{Azzam77,Roseler90} (In the following we will discuss experimental situations in reflection geometry
although most of the considerations are valid for transmission, as well.) A representative example
is the so-called rotating-analyzer ellipsometry (RAE) when the reflectivity is measured at a finite angle of
incidence, usually in the vicinity of the Brewster angle.\cite{Azzam77,Roseler90} Under this condition the
Fresnel coefficients are different for polarization parallel ($p$-wave) and perpendicular ($s$-wave) to the
plane of incidence and the initially linearly polarized light becomes elliptically polarized upon the
reflection. By rotating the analyzer the ellipsometric parameters, i.e. the phase difference and the intensity
ratio for the $p$-wave and $s$-wave components of the reflected light,\cite{note} are measured and the complex
refractive index can be directly obtained. Each of the above experimental methods is far more complicated than
the measurement of unpolarized reflectivity or transmittance near normal incidence.

As an alternative, we describe a new type of ellipsometry, hereafter referred to as double-reference
spectroscopy (DRS). It offers a simple way to obtain the complex dielectric function without KK transformation
by measuring the unpolarized normal-incidence reflectivity of the sample relative to two transparent reference
media. In addition to the absolute reflectivity, i.e. that of the vacuum-sample interface, we can take advantage
from the excellent optical properties of diamond and use it as a second reference. High-quality such as type IIA
optical diamonds are transparent from the far-infrared up to the ultraviolet photon energy
region,\cite{Edwards85} except for the multiphonon absorption bands located at
$\omega=0.19-0.34$\,eV.\cite{Thomas95} Moreover, they have a large refractive index $n_d\approx2.4$ which shows
only $10\%$ energy dependence up to $\omega \approx
5$\,eV.\cite{Philipp64,Edwards85,Edwards81,Thomas95,Djurisic98} The significant difference between the
refractive index of the two reference media is a crucial point of the method.

As the experimental arrangement is much simpler than that of any ellipsometric technique, this method may find a
much broader application, especially when only a narrow spectral range is of interest (or when Kramers-Kronig
transformation cannot be performed due to experimental limitations). A representative field of application is
the class of strongly correlated electron systems when the optical properties are very sensitive to the low
energy excitations. In this case the KK transformation cannot reliably differentiate between full- or pseudo-gap
behavior or cannot even distinguish between bad metals with fully incoherent low-frequency response and
semiconductors with small charge gap.
\begin{figure}[th!]
\includegraphics[width=3.4in]{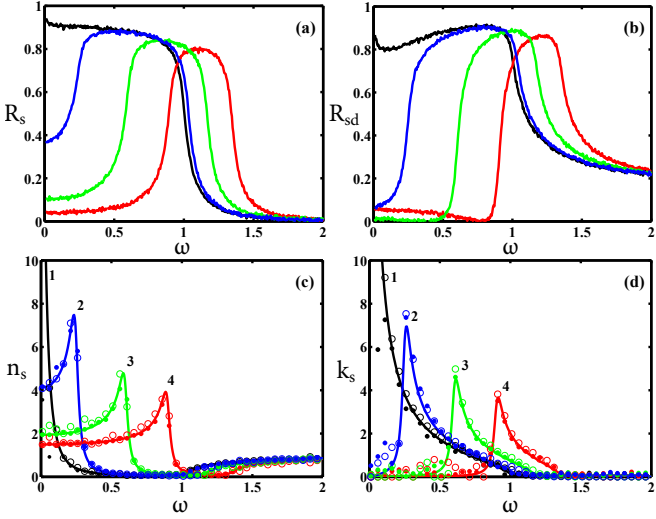}
\caption{(Color online) Panel (a)-(b): Normal-incidence reflectivity spectra of the vacuum-sample ($R_{vs}$) and diamond-sample ($R_{ds}$) interfaces as calculated from the dielectric function $\hat{\epsilon}_s(\omega)=1+(\omega_0^2-\omega^2-i\omega\gamma)$ for $\omega_0=0, 0.25, 0.6, 0.9$ and $\gamma=0.1$. The unit of the energy scales corresponds to the plasma frequency. Gaussian noise with $\Delta R=\pm0.005$ standard deviation is introduced to both $R_{vs}(\omega)$ and $R_{ds}(\omega)$. Panel (c)-(d): The refractive index ($n_s$) and the extinction coefficient ($k_s$) as obtained from the above reflectivities using the DRS (closed circles) and the RAE (open circles) approach. The complex refractive index free of noise is indicated by full lines.}
\end{figure}

In the following, we describe the principles of the double-reference spectroscopy and demonstrate its efficiency
in comparison with the RAE. The essence of the method is the measurement of the sample reflectivity relative to
two media with strongly different dielectric properties, such as vacuum and diamond. For normal incidence the
Fresnel equations for the two interfaces have the following form:
\begin{eqnarray*}
R_{vs}\equiv R_s=\left|\frac{\hat{n}_s-1}{\hat{n}_s+1}\right|^2\ \ \
\ \ and\ \ \ \ \
R_{ds}=\left|\frac{\hat{n}_s-n_d}{\hat{n}_s+n_d}\right|^2\ ,
\end{eqnarray*}
where $\hat{n}_s=n_s+ik_s$ denote the complex index of refraction and $\hat{n}_d(\omega)$ is well documented in
the literature for type IIA diamonds.\cite{Philipp64,Edwards81,Edwards85,Djurisic98,Thomas95} Although the
Fresnel equations are highly nonlinear, $\hat{n}_s$ can easily be expressed in the lack of absorption within the
diamond, i.e. for $k_d\equiv0$:
\begin{eqnarray}
n_s&&=\frac{1}{2}(n_d^2-1)\left(n_d\frac{1+R_{ds}}{1-R_{ds}}-\frac{1+R_{s}}{1-R_{s}}\right)^{-1}\ ,\\
k_s&&=\left(-n_s^2+2\frac{1+R_{s}}{1-R_{s}}n_s-1\right)^{1/2}\ .
\end{eqnarray}

We show the efficiency of the method using the model dielectric function
$\hat{\epsilon}_s(\omega)=1+(\omega_0^2-\omega^2-i\omega\gamma)^{-1}$, where $\omega_0$ and $\gamma$ are the
resonance frequency and the damping of the oscillator, respectively. From $\hat{\epsilon}_s(\omega)$ we evaluate
both $R_s(\omega)$ and $R_{ds}(\omega)$ by the Fresnel equations while in case of RAE the intensity is
calculated for three different orientations of the analyzer.\cite{note} The resonance frequency is varied in
such a way that the reflectivity spectra, plotted in the upper panels of Fig.~1, describe both insulating and
metallic behavior, corresponding to $\omega_0>0$ and $\omega_0=0$, respectively. The typical noise of the
detection and the finite energy resolution are taken into account as Gaussian noise superimposed on the
intensities with standard deviation of $\Delta R=\pm0.005$. Furthermore, systematic errors coming from the
imperfect experimental conditions -- such as sample surface roughness and non-planarity, misalignment of the
light path -- are represented as $1^{\circ}$ deviation in the angle of incidence for the both cases.
\begin{figure}[th!]
\includegraphics[width=3.2in]{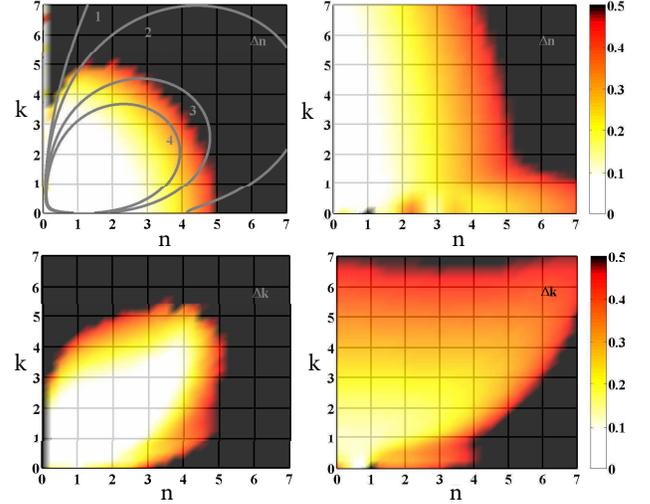}
\caption{(Color online) Color map of the absolute error of the complex refractive index as calculated by the DRS (left panels) and the RAE (right panels). The upper and lower panels show the error for the refractive index ($\Delta n_s$) and the extinction coefficient ($\Delta k_s$), respectively. Full lines with labels correspond to the spectra shown in Fig.~1.}
\end{figure}

From the given reflectivity spectra the complex refractive index is calculated by the double-reference method
using Eqs.~1-2 and also by following the more complicated evaluation of RAE. The real and imaginary part of the
respective $\hat{n}_s(\omega)$ spectra are shown in the lower panels of Fig.~1. The exact
spectra, $\hat{n}_s(\omega)=\sqrt{\hat{\epsilon}_s(\omega)}$, calculated without introducing experimental errors
are also shown for comparison. The precision of the two methods seems comparable.

In order to classify the range of applicability, the error maps for the two techniques are analyzed in more
detail over the plane of the complex refractive index. As Fig.~2 shows, the overall confidence level of the DRS
surpass that of the RAE, especially in case of the extinction coefficient, $k_s$. Furthermore, while the error
map for the real and imaginary part of the refractive index behaves similarly in case of the DRS, the RAE is
optimal for the two components in rather distinct regions of the $n_s-k_s$ plane. Although for RAE the area of
applicability is seemingly more extended for the real part of the refractive index in the limit of $k_s\gg n_s$, the extinction coefficient dominating the optical response in this strongly-absorbing region exhibits a high error level.
\begin{figure}[th!]
\includegraphics[width=2.4in]{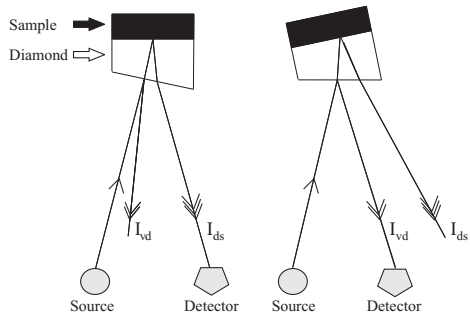}
\caption{Experimental condition for the measurement of $R_{ds}$.
Reflection from the vacuum-diamond and diamond-sample interfaces are indicated.
Multiple reflections within the diamond can be neglected due to the wedging of the
window. This wedging also allows for a clean separation of the reflections from
the two interfaces and thus facilitates reference measurements (see text for details).}
\end{figure}

In situations where $\mid$$\hat{n}_s$$\mid\gg$$1$, such as strong resonances or good metals with large extinction coefficient ($k_s$$>$$n_s$$\gg$$1$), the Fresnel equations are not numerically
independent since the reflection coefficients converge to the unity irrespective of the polarization state of
the light, the angle of incidence or the reference media. In this limit, the both approaches fail which is
general for any ellipsometry. For the DRS it means that the difference between
the two reference media disappears as $\mid$$\hat{n}_s$$\mid\gg$$n_d$. With a realistic noise level specified above, the DRS works with less than
$\sim10\%$ error until the refractive indexes are twice as large as that of the diamond, i.e. almost in the
whole range of $\mid$$\hat{n}_s$$\mid\leq5$. It is to be emphasized at this point that the large
difference in the refractive index of the two reference media highly extends the applicability range of
the method and reduces the numerical errors.

The unique efficiency of diamond arises from its wide transparency window. Note, however, that optically
well-characterized semiconductors, such as Si,\cite{Li80} GaAs,\cite{Skauli03} and CdTe,\cite{Hlidek01} can
provide an even better better performance for a limited range of energy, typically below $\omega\approx1$\,eV.
Since these materials are popular substrates for crystal growing, DRS can be carried out by the successive
measurement of the two sides of the samples.

Next we describe a simple procedure for the measurement of $R_{ds}$ applying a wedged diamond piece as sketched
in Fig.~3.\cite{Kezsmarki07} The intensity reflected back from the vacuum-diamond and diamond-sample interfaces ($I_{vd}$ and $I_{ds}$, respectively) can be detected separately by a few degree rotation; a wedging angle
of $2^{\circ}$ causes $\sim10^{\circ}$ angular deviation between the two reflected beams. Since the nearly normal
incidence can still be considered for both positions, the reflectivity of the sample relative to the diamond
is obtained from the measured intensities as:\cite{Kezsmarki07}
\begin{equation}
R_{ds}(\omega)=\frac{R_{vd}(\omega)}{(1-R_{vd}(\omega))^2}\cdot\frac{I_{ds}(\omega)}{I_{vd}(\omega)}\,
\label{eq1}
\end{equation}
where $R_{vd}$ ($\equiv R_d$) is the absolute reflectivity of the
diamond. The $R_{vd}(\omega)/(1-R_{vd}(\omega))^2$ prefactor can be either
calculated from the well-documented refractive index of diamond,
\cite{Philipp64,Edwards81,Edwards85,Djurisic98,Thomas95} or checked experimentally.

The high-energy limit of this method is mainly determined by the roughness of the diamond-sample interface
$\delta_{ds}$. Therefore, special care should be taken for the proper matching between the diamond and the sample
in order to eliminate interference and diffraction effects inherently appearing for wavelength shorter than
$\delta_{ds}$.

In conclusion, we have described a new ellipsometric method whose applicability is demonstrated both
for insulating and metallic compounds. The experimental performance, which is far more simple as
compared with other ellipsometric techniques, means the measurement of the
normal incidence reflectivity relative to two reference media, e.g. the reflection from the vacuum-sample and diamond-sample interfaces. The double-reference method may find broad application either in the field of broadband optical spectroscopy or in material characterization due to its numerical precision and simplicity.

\section*{Acknowledgement}
The authors are grateful to L. Forr\'o, R. Ga\'al, G. Mih\'aly and L. Mih\'aly for useful discussions. This work
was supported by the Hungarian Scientific Research Funds OTKA under grant Nos. F61413 and K62441 and Bolyai 00239/04.

%
%

%
\end{document}